\documentclass[12pt]{article}
\usepackage{graphics}
\input epsf

\newcommand{\be}{\begin{eqnarray}}
\newcommand{\ee}{\end{eqnarray}}

\begin{document}
\title{
\begin{flushright}
{\large UWThPh-2003-20}\\
{\large UAHEP03010}
\end{flushright}
\vskip 1cm
R Parity Violating Decays of the Gluino}
\author{L. Clavelli\footnote{lclavell@bama.ua.edu}\\
Department of Physics and Astronomy, University of Alabama,\\
Tuscaloosa AL 35487\\ \and H.
Stremnitzer\footnote{hanns.stremnitzer@univie.ac.at}\\ Institut
f\"ur Theoretische Physik, Universit\"at Wien, A-1090 Vienna,
Austria\\} 
\date{August 26, 2003}
\maketitle
\begin{abstract}
Assuming the lightest supersymmetric particle is the gluino,
we treat the 
decays ${\tilde g} \rightarrow q {\overline q} \nu$ and
${\tilde g} \rightarrow g \nu$.  Such couplings can be induced
by the R parity violating quark-squark-lepton interaction which
can also be responsible for neutrino masses and mixings.  These
R parity violating gluino decays have the same final state
structure (jets plus missing energy) as previously considered
decays into quark-antiquark-photino and gluon-gravitino but with
significantly different gluino lifetimes.
\end{abstract}
%\pacs{PACS numbers: 11.30.Pb, 12.60.J, 13.85.-t}
\renewcommand{\theequation}{\thesection.\arabic{equation}}
\renewcommand{\thesection}{\arabic{section}}
\section{\bf Introduction}
\setcounter{equation}{0}

   The possibility of the lightest supersymmetric particle (LSP)
decaying through R parity violation has been discussed since the
early days of supersymmetry \cite{HaberKane}, many studies have
been made (e.g. \cite{Herz}), and many experimental searches have
been undertaken (e.g. \cite{Aleph02}).  There seem, however, to
have been no investigations up to now of the possibility
considered here of R parity violating decays of the gluino.
Although these would be most relevant if the recent theme of a
light or relatively light gluino were to be experimentally
confirmed they could also be important in the case of a heavy
gluino LSP. In some cases \cite{Bartl,CGM} the phenomenological
advantages of a light gluino have as much to do with a large
hierarchy between gluino and squark masses as with a light gluino
mass per se. Most recent light gluino investigations have
concentrated on the mass region around $12$ GeV possibly
accompanied by a b squark near $4$ GeV \cite{BHKSTW} , \cite{LG}.
Alternative possibilities have also been noted \cite{newposs}. In
the minimal supersymmetric standard model (MSSM), a light gluino
is typically accompanied by an even lighter photino allowing the
decay ${\tilde g} \rightarrow q {\overline q} {\tilde \gamma}$. In
gauge mediated SUSY breaking models, there is often an ultra-light
gravitino below the gluino in mass leading to the decay ${\tilde
g} \rightarrow g {\tilde G}$.  In some models these channels are
closed leading to an absolutely stable gluino or to one decaying
through R parity violating processes. In this article we assume
the gluino is the lightest supersymmetric particle and we consider
lepton number violating gluino decays.  In section II we treat the
decays ${\tilde g} \rightarrow q {\overline q} \nu$. The fact that
the third generation seems, in some respects, special compared to
the first two suggests models in which the R parity violation is
tied to the third generation.

If the R parity violation is entirely in the third generation
and the gluino lies below the $b {\overline b}$ threshold, there will
be the loop induced decay ${\tilde g} \rightarrow g \nu$.  This is
treated in section III neglecting the possibility of a right-handed
neutrino.

Finally, in the presence of a light right-handed neutrino, there
is the possibility, treated in section IV, that the dominant decay
could be a two loop process coupling the gluino to gluon plus
$\nu_{R}$. It might be expected that such a dominant decay
mechanism would lead to an ultra-long-lived gluino.

\section{\bf The gluino decay to quark-antiquark-neutrino}
\setcounter{equation}{0}

      We assume an $R$ parity violating and lepton number violating term in
the superpotential of the form
\be
     W = \lambda ^\prime_{ijk} L_i Q_j D_k
\label{RPVterm} \ee where $i,j,k$ are family indices. Then, if the
gluino is above a quark-antiquark threshold one will have the
gluino decay to $q \overline{q} \nu$ corresponding to the graph of
fig. 1.

\par

\begin{figure}[tb]
\begin{center}
%\begin{picture}(470,160)(0,0)
\begin{picture}(470,260)(0,0)
%\graphpaper[10](0,0)(480,130)
\put(-5, 5){\mbox{\resizebox{320pt}{!}{\includegraphics{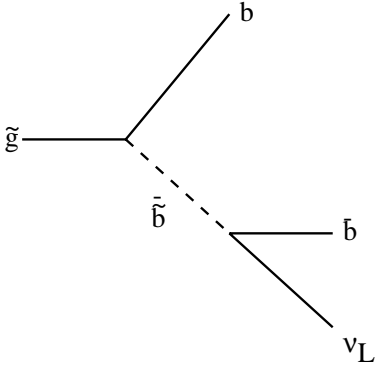}}}}
% 480pt
\end{picture}
\end{center}
%\vspace*{-1cm}
\vspace*{-3cm}
\caption{Gluino decay into quark-antiquark neutrino}
\label{fig1}
\end{figure}

The fact that the third generation seems, in some respects,
special compared to the first two suggests models in which the R
parity violation involves third generation quarks and squarks
only. We, therefore, take as a working assumption that the
non-zero quark flavor indices in \ref{RPVterm} are third
generation only. In any case, the experimental limits on R parity
violating couplings are much less restrictive \cite{Herz,Raychaud}
in the third generation so these could well be dominant. A
$\lambda^\prime$ involving only third generation quarks could be
as great as $0.1$. Then the gluino decay would be into $b
{\overline b} \nu$ assuming the gluino mass is above the $2 b$
threshold.  In this case a light gluino pair production could
explain the excess b production seen at Fermilab without requiring
the light b squarks of \cite{BHKSTW}.

The signature of such a decay, hadrons plus missing energy, would be
identical to the conventional $q \overline{q} \tilde \gamma$ decay.
The inverse width for this latter decay is
\be
     \tau(\tilde g \rightarrow q \overline q {\tilde \gamma}) \approx
   {\tilde m}^4 /(m_{\tilde g}^5 \alpha_s \alpha)
\ee With a light gluino and a squark mass in the $100 GeV$ range,
this typically results in a gluino lifetime in the nanosecond
range which is counter-indicated by the $KTEV$ search \cite{KTEV}.
If, however, the gluino is the LSP and decays through the R parity
violating graph of fig. 1, the lifetime could easily be much
longer so that the lightest supersymmetric hadron, presumably the
gluino-gluon bound state, would not decay in the sensitive region
of a $KTEV$ type experiment.
\footnote{We have recently become aware of a previous treatment
\cite{Dreiner}of this decay mode.} 

\be
     \tau(\tilde g \rightarrow q \overline q \nu) \approx
   {\tilde m}^4 /(m_{\tilde g}^5 \alpha_s {\lambda ^\prime}^2)
\label{gluinolifetime} \ee 
The neutrino mass matrix corresponding
to the R parity violation of \ref{RPVterm} is \cite{Barbier}
\be
     M^{\nu}_{ii^\prime} \approx \frac{3}{8 \pi^2} \sum_{jk}
      \lambda^\prime_{ijk}\lambda^\prime_{ikj}m_d(j)m_d(k)/{\tilde m}
\ee
where $\tilde m$ is the assumed degenerate squark mass and
$m_d(k)$ is the down-type quark mass in the k'th family.
We have also made the simplification of
neglecting the CKM quark mixing angles and have assumed that
$A_d - \mu \tan \beta$ is of order $\tilde m$.  If the dominant
$k,j$ are third generation, \cite{Barbier} notes that a $4.5 eV$
$\nu_e$ mass would correspond to
\be
     \lambda^\prime_{133} \approx 7 \cdot 10^{-4} \left(
     \frac{\tilde m}{100 GeV} \right )^{1/2}
\label{limit}
\ee
although the present indications of a sub-eV neutrino mass would
lead to a $\lambda^\prime$ an order of magnitude below \ref{limit}.
Such a relation substituted into \ref{gluinolifetime} would lead to
a gluino lifetime depending only on the ratio of gluino mass to
squark mass.
\be
  \tau(\tilde g) = .013 s \left( \frac{\tilde m}{1000 m_{\tilde g}}
    \right)^5 \left ( \frac{0.1}{\alpha_s} \right )
    \left ( \frac{7 \cdot 10^{-5}}{\lambda^\prime_{i33}} \right )^2
\ee
A gluino-gluon bound state with a lifetime of order $.013s$ might
have evaded the current searches since it would appear as a
quasi-stable particle whose ultimate decay with a missing neutrino
could be confused with a neutron knock-on event.

\section{\bf The gluino decay to gluon plus neutrino}
\setcounter{equation}{0}

\begin{figure}[tb]
\begin{center}
%\begin{picture}(470,200)(0,0)
%\begin{picture}(432,360)(0,0)
\begin{picture}(432,300)(0,100)
%  bounding box = 54 360 486 720   432=486-54   360=720-360
%\graphpaper[10](0,0)(480,190)
\put(-50,5){\mbox{\resizebox{480pt}{!}{\includegraphics{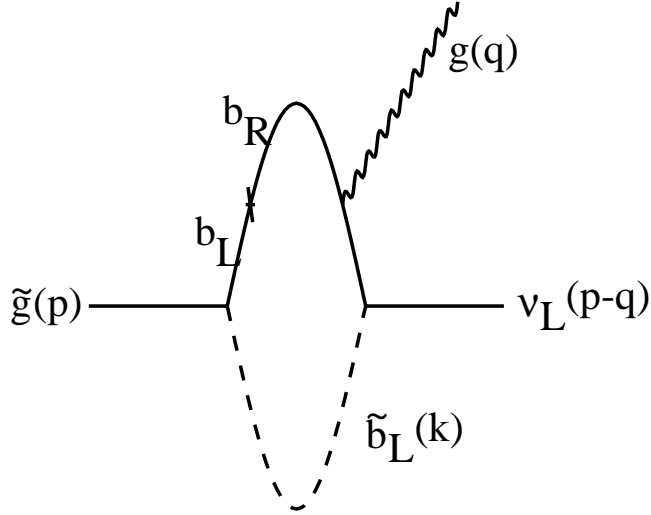}}}}
\end{picture}
\end{center}
%\vspace*{-1cm}
\caption{Gluino decay to gluon plus left-handed neutrino}
\label{fig2}
\vspace*{2cm}
\end{figure}

If the gluino mass is less than twice the mass of the quarks
appearing in the $R$ parity violating couplings, one has the
hitherto uninvestigated decay
\be
     {\tilde g} \rightarrow g \nu  .
\ee
A recent analyses of constraints from Z decay suggests at
95\% confidence level \cite{Aleph97,Janot}
\be
     m_{\tilde g} > 6.3 GeV/c^2  .
\ee

This limit still allows the possibility that the gluino mass is
below the b quark pair threshold.  The matrix element
corresponding to the graph of fig. 2 plus the analogous graph
where the gluon is emitted from the squark line is
%\newpage
\be
\nonumber
  {\cal M} = \epsilon_{\mu} \frac{\lambda ' {g_3}^2 m_b}{4} {\overline u}(p')
  \int \frac{d^4 k}{(2 \pi)^4} \left ( \frac{(\not p' + \not k)\gamma^\mu
   + \gamma^\mu (\not p + \not k)}{(p+k)^2 - m_b^2} + \frac{(2 k - q)^\mu}
   {(k-q)^2 - {\tilde m}^2} \right ) \\
   \cdot (1-\gamma_5) u(p) \frac{1}{(p'+k)^2-m_b^2}
   \frac{1}{k^2-{\tilde m}^2}
\ee
where $p'=p-q$ is the final state neutrino momentum.
\vskip 1cm
\begin{figure}[h]
\begin{center}
%\begin{picture}(470,200)(0,0)
%\begin{picture}(432,360)(0,0)
\begin{picture}(360,188)(0,100)
%  bounding box = 54 432 414 720   414-54=360  720-432=288
%  bounding box = 54 360 486 720   432=486-54   360=720-360
%\graphpaper[10](0,0)(480,190)
\put(-100,5){\mbox{\resizebox{480pt}{!}{\includegraphics{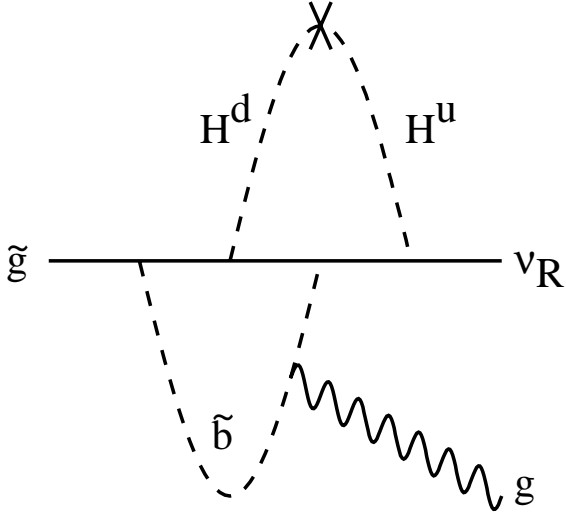}}}}
\end{picture}
\end{center}
%\vspace*{-1cm}
\caption{Typical Feynman graph for gluino decay to gluon plus
right-handed neutrino}
\label{fig3}
\vspace*{2cm}
\end{figure}

This amplitude is equivalent to that induced by an effective
magnetic moment coupling
\be
   {\cal L} = i \mu_a {\overline u}_L(\nu)\sigma^{\mu \nu} q_\nu
     u ({\tilde g})\epsilon_\mu(g)
\ee
with
\be
      \mu_a \approx \frac{m_b \alpha_s \lambda^\prime}{{\tilde m}^2}
\label{mua1}
\ee
The corresponding decay rate is
\be
   \Gamma({\tilde g} \rightarrow g \nu) = \mid \mu_a \mid ^2 m_{\tilde
g}^3 \approx \frac{m_b^2 m_{\tilde g}^3 \alpha_s^2 \lambda^{{\prime}2}}
{{\tilde m}^4}
\label{gammag}
\ee
Nominal values of $10 GeV$, $10 TeV$, $0.1$, and $7 \cdot 10^{-5}$ for $m_{\tilde g}$, $\tilde m$, $\alpha_s$, and $\lambda^\prime$,
would correspond to a gluino lifetime of $5.2 \cdot 10^{-3} s$.

     A two loop graph that would lead to a gluino decay to gluon
plus right-handed neutrino is shown in fig. 3.  To investigate
such two loop effects we consider an extended superpotential
containing a right-handed singlet superfield N.
\be
     W = W_{MSSM}+ \lambda^\prime Q D L + \epsilon H^u H^d N + h_\nu L N H^u
\label{superpot}
\ee

   Here, lepton number and R parity violation comes through
a new $H^u H^d N$ interaction governed by the coupling constant
$\epsilon$ in addition to the conventional $\lambda^\prime$
coupling. $W_{MSSM}$ contains the usual Higgs mixing term $\mu H^u
H^d$. The small Yukawa coupling $h_\nu$ is proportional to the
neutrino Dirac mass:
\be
       h_\nu = \frac{m_\nu^{Dirac}}{<H^u>}
\ee

    The effective transition magnetic moment from fig. 3 is
\be
     \mu_a \approx \frac{1}{(4 \pi)^3}
         \frac{ \lambda^\prime h_\nu h_b \alpha_s \mu B m_b}
               {{\tilde m}^2 m_H^2}
\ee
    Here, $\mu B$ is the off-diagonal entry in the
Higgs mass squared matrix. In the MSSM with electroweak symetry
breaking \cite{Castano}, it is given by $\mu B = \frac{1}{2}M_A^2
sin (2\beta)$ where $M_A$ is the mass of the CP--odd Higgs scalar.

This $\mu_a$, however,
would be proportional to the neutrino mass and would be negligible
compared to the transition magnetic moment of \ref{mua1}.

\section{\bf A dominant gluino decay to gluon plus right-handed neutrino}
\setcounter{equation}{0}

     Even in the absence of the R parity violating quark-squark-lepton
coupling of \ref{RPVterm}, the R parity and
lepton number violating Higgs-Higgsino-Lepton coupling in \ref{superpot}
could lead to a gluino decay into gluon plus right-handed neutrino.

\begin{figure}
\begin{center}
\includegraphics{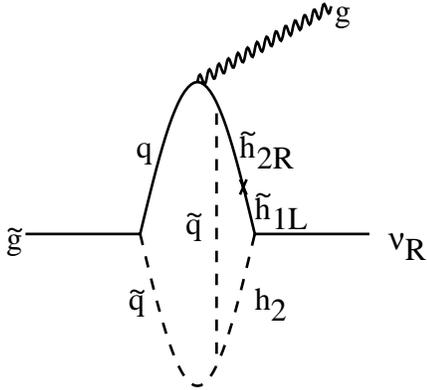}
\end{center}
\vspace{-2cm}
\caption{Alternative mechanism for gluino decay into gluon plus
right handed neutrino}
\label{fig4}
\end{figure}

The lowest order graph contributing to gluino decay would be that
of fig. 4 as well as those graphs related by attaching the gluon
to other colored lines or by changing the flavor of the internal
quark (squark) lines.  These amplitudes are proportional to the
trilinear boson coupling parameters $A$, which are induced in the
softly broken MSSM thru supergravity. We can entertain the
possibility that the $\lambda^\prime$ parameters are negligible
and that the dominant trilinear coupling is that of the top quark.
Then the two loop gluino decays of fig. 4 could be dominant for a
gluino LSP of mass up to the minimum of twice the top mass or
twice the stop mass:
\be
    \mu_a \approx \frac{1}{(4 \pi)^3} \frac{ \mu \cos(2 \beta)
    m_{\chi^0} A \epsilon \alpha_s h_t}{{\tilde m}^2 m_H^2}
\ee

where $m_{\chi^0}$ is the mass of the neutralino.  However, if the
top quark is heavier than the neutralino, then $m_{\chi^0}$ should
be replaced by $m_t$ in the estimate here.

 Depending on the
flavor structure, the $\epsilon$ coupling might be limited only by
neutrino mass measurements and might, therefore, be expected to be
extremely small. It is clear that the corresponding gluino
lifetime could easily be long enough to have cosmological
significance. There is, for instance, the possibility that the
gluino-gluon bound state could traverse cosmological distance
scales before decaying and, if sufficiently energetic, could
contribute through its decay products to the ultra-high-energy
cosmic rays.  We leave delayed calculation of some of the possible
effects discussed here to future investigations.

{\bf Acknowledgements}

    This work was supported in part by the US Department of Energy
under grant DE-FG02-96ER-40967.  One of the authors, (LC), would
like to acknowledge the hospitality of the University of Vienna
and the High Energy Physics Institute of the Austrian Academy of
Sciences where the initial discussion of the ideas presented here
occurred. Discussions with Alfred Bartl as well as some
conversations on the gluino decay modes of the current paper with
Thomas Gajdosik are especially acknowledged.  It has been brought
to our attention that gluino decays through bilinear R parity
violation (as opposed to the trilinear violation treated here)
have also been recently considered \cite{HirschPorod}.

\par

\end{document}